\documentclass[aps,prb,twocolumn,amsmath,amssymb,showpacs,superscriptaddress,notitlepage,longbibliography]{revtex4-2}

\usepackage[colorlinks=true,linkcolor=blue,anchorcolor=red,citecolor=blue, urlcolor=blue]{hyperref}
\usepackage{bm}
\usepackage{graphicx}
\usepackage{xcolor}
\usepackage{verbatim}

\begin{document}

\title{Thermoelectric transport of the coexistence topological semimetal in the quantum limit}

\author{L. W. Guo}
\author{C. M. Wang}
\email{Corresponding author: wangcm@shnu.edu.cn}
\affiliation{Department of Physics, Shanghai Normal University, Shanghai 200234, China}

\begin{abstract}
We explore the thermoelectric transport properties of a coexistence topological semimetal, characterized by the presence of both a pair of Weyl points and a nodal ring in the quantum limit. This system gives rise to complex Landau bands when subjected to a magnetic field aligned with the direction connecting two Weyl points. In the longitudinal configuration, where the magnetic field is parallel to the electric field or the temperature gradient, the thermoelectric conductivity indicates a plateau independent of the magnetic field and the Fermi energy at $\delta$-form short-range scattering. This platform structure should also exist in pure two-node Weyl semimetals. However, the thermoelectric conductivity and the Seebeck coefficient are significantly influenced by the parameters of long-ranged Gaussian or screened Coulomb scattering potentials for both fixed carrier density and Fermi energy scenarios. In the transverse configuration, both Gaussian and screened Coulomb scatterings yield substantial positive magnetoresistance and thermoelectric conductance. Since the Hall conductivity is larger than the longitudinal one, the Seebeck coefficient, exhibiting a quadratic increase with the magnetic field, is close to the dissipationless limit irrespective of scatterings, while the Nernst response is notably dependent on the scattering mechanism. Additionally, the model parameter, distinct from the two-node Weyl model, influences the thermoelectric transport properties. The magnetic field response of the thermoelectric coefficients to different scattering potentials can be used as a basis for distinguishing scattering mechanisms in materials.
\end{abstract}

\date{\today}
\maketitle

\section{Introduction}
Topological semimetals \cite{Wan11prb,Xu11prl,Burkov11prl,Weng15prx,Lv15prx,Armitage18rmp,Lv2021rmp} have recently attracted increasing attention due to their non-trivial and exotic physical phenomena: their band structure includes symmetry-protected band crossings between the valence and conduction bands, and they exhibit unconventional responses to applied external fields \cite{Zyuzin2012prb,Aji12prbrc,WangCM16prl,WangCM17prl,Zhang2019prl}. Particularly, as thermoelectric effects are proportional to the derivative of conductivity, they are generally considered more sensitive to these novel responses \cite{Liang13nc, Hirschberger16nm, Jia16nc, Liang17prl, Matusiak17nc, Zhang2021berry, Alam2023prb,Zhang2021prb}. In recent years, theoretical studies of three-dimensional Dirac/Weyl semimetals in the quantum limit, where only the lowest Landau band is occupied by electrons, have shown that the Seebeck coefficient $S_{xx}$ indicates an unsaturated linear growth at low temperatures \cite{Skinner18sa}. Further investigation demonstrated that the linear growth of the Seebeck coefficient in the clean limit originates from the the plateau value of thermoelectric Hall conductivity $\alpha_{xy}$ \cite{Kozii2019prb,Pratama2022prb}, proportional to the temperature. The plateau of $\alpha_{xy}$ or the equivalent linear growth of $S_{xx}$, recognized as a signature of topological semimetals, has been observed in the Dirac semimetal ZrTe$_5$ \cite{Zhang2020nc, Wang2021prb}, Pb$_{1-x}$Sn$_x$Se \cite{Liang13nc}, Weyl semimetal TaP \cite{Han2020nc}, and nodal-line semimetal graphite \cite{Kiswandhi2023prb, Osada2022jpsj}. Hence, topological semimetals show great potential in the realm of thermoelectric transport.

Recently, a distinctive class of topological states, termed ``topological coexistence'', has been identified in certain topological materials \cite{Zhao2021prb, zhang2021prr, Li2022prb, Wu2023apl, Zhan2023prb, saini2023arxiv, Chang2016SR,sun2017prb}. Diverging from traditional topological semimetals characterized by a single topological state, these materials showcase the simultaneous presence of multiple topological states in the bulk. One notable example is found in the double perovskites Ba$_2$CdReO$_6$ \cite{Zhao2021prb}, which host both Weyl and nodal ring topological states. In a related study, Zhang et al. \cite{zhang2021prr} investigated the surface states of the ferromagnetic material Cs$_2$MoCl$_6$, sharing the same space group $Fm\bar{3}m$ as Ba$_2$CdReO$_6$ and featuring analogous topological coexistence states. Their findings reveal a unique surface state connection between Weyl and nodal ring fermions, a phenomenon also predicted in other systems \cite{Wu2023apl, Zhan2023prb}.

Topological coexisting materials, as a new class of topological phases, have energy bands that are quite different from those previously studied, which may bring special thermoelectric transport properties. In the quantum limit, there is only one energy band that contributes to transport. At this point, scattering may play a crucial role \cite{ZhangSB16njp,ji2017competition,Konye2019prb,feng2020jpcm,Fu2022prb,LiY2023prb,Li2023prb}. Our study starts from a two-band model containing both a pair of Weyl points and a nodal ring, and we investigate the longitudinal as well as transverse thermoelectric transport properties in the quantum limit when the magnetic field is applied along the connection of Weyl points by considering both the Gaussian and screened Coulomb electron-impurity scatterings. 

\section{Hamiltonian and the Landau bands}\label{model}
\begin{figure}[tbp]
\centering
\includegraphics[width=1\columnwidth]{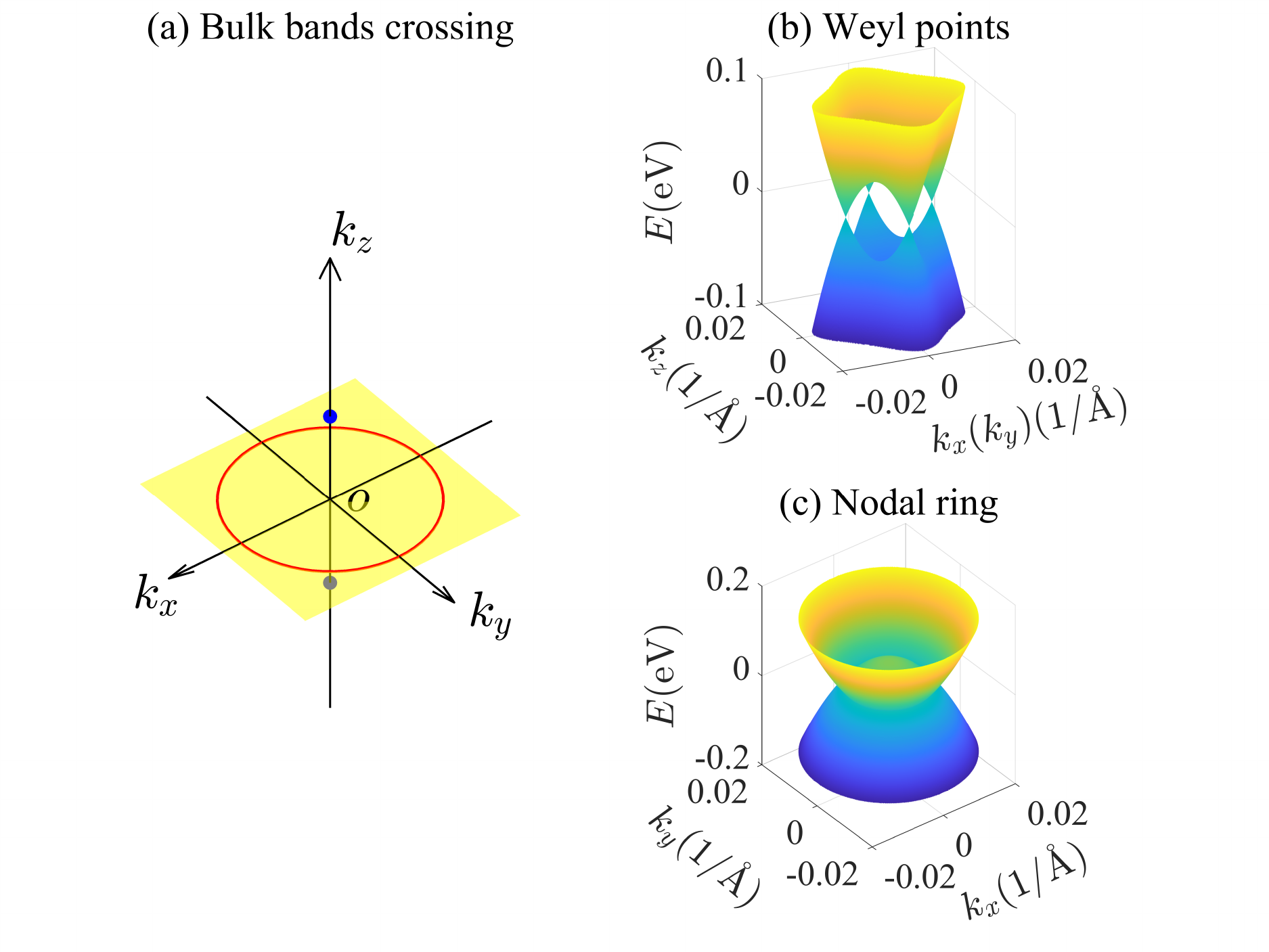}
\caption{(a) The schematic diagram of the band crossing in momentum space, which forms a pair of Weyl points (blue and gray dots) and a nodal ring (red circle). The yellow plane represents the $k_z=0$ plane. Energy dispersions in the $k_xOk_z$ or $k_yOk_z$ plane (b) and the $k_xOk_y$ plane (c). Here $M_0=0.05$ eV, $M_1=5$ eV nm$^2$, and $D=2M_1$.}
\label{fig-dispersion}
\end{figure}

We consider a topological semimetal, where a pair of Weyl nodes and a nodal line coexist. The minimal Hamiltonian is written as \cite{zhang2021prr}
\begin{align}
	H=Dk_z(k_x\sigma_x+k_y\sigma_y)+M_{\bm k}\sigma_z,
\end{align}
with
\begin{align}
	M_{\bm k}=M_0-M_1(k_x^2+k_y^2+k_z^2).
\end{align}
Here $D$, $M_0$, and $M_1$ are material parameters, $\bm \sigma=(\sigma_x,\sigma_y,\sigma_z)$ are three $2\times2$ Pauli matrices, and $\bm k=(k_x,k_y,k_z)$ is the wave vector.
The dispersion of the system is
\begin{align}
	E_{\lambda\bm k}=\lambda\sqrt{D^2k_z^2(k_x^2+k_y^2)+M_{\bm k}^2}.
\end{align}
The quantum number $\lambda=\pm1$ denotes the conduction and valence bands. Compared to the minimal two-node Weyl semimetal model \cite{Lu15Weyl-shortrange} $H_W=A(k_x\sigma_x+k_y\sigma_y)+M_{\bm k}\sigma_z$, the constant Fermi velocity $A$ becomes a momentum-dependent quantity $Dk_z$, which results in the coexistence of two types of topological states. Two bands contact each other at a pair of Weyl points ($0,0,\pm\sqrt{M_0/M_1}=\pm k_W$) on the $k_z$ axis, in addition to a closed nodal line (nodal ring) in the $x$-$y$ plane ($k_z=0$, $k_x^2+k_y^2=k_W^2$), displaying two topological properties in one system. Figure \ref{fig-dispersion}(a) schematically shows the characteristics of band crossing formed by two energy bands of the model, where blue and gray dots represent a pair of Weyl points on the $k_z$ axis, and the red circle represents the nodal ring in the $k_x$-$k_y$ plane. The panel (b) shows the energy dispersion on the $k_xOk_z$ or $k_yOk_z$ plane, where two additional Weyl points come from the intersections of the ring and the $k_x$ or $k_y$ axis. The panel (c) indicates the dispersion in the $k_z=0$ plane, where the touch points compose a closed circle. 

The Berry curvature is considered to be an important quantity reflecting the topological properties of matter. It can be thought of as the magnetic field in momentum space and is associated with various topological transport responses such as the anomalous Hall effect, the quantum Hall effect and the chiral anomaly. Three components of the Berry curvature for the coexistence model can be found as:
\begin{align}
    \Omega_+^x =& \frac{D^2}{2E_+^3}k_zk_x[M_0-M_1(k_x^2+k_y^2-k_z^2)],\\
    \Omega_+^y =& \frac{D^2}{2E_+^3}k_zk_y[M_0-M_1(k_x^2+k_y^2-k_z^2)],\\
    \Omega_+^z =& -\frac{D^2}{2E_+^3}k_z^2[M_0+M_1(k_x^2+k_y^2-k_z^2)].
\end{align}
Here the ``+" subscript represents the Berry curvature of the $\lambda=+1$ band. In contrast to the two-node Weyl semimetal \cite{Lu15Weyl-shortrange}, three components of the Berry curvature all take zero in the $k_z=0$ plane. By calculating the integral of the Berry curvature around the Fermi surface near the $\pm k_W$ point, we can obtain a topological charge $n=\pm 1$, corresponding to a pair of Weyl points with opposite chiralities. In addition, we can define a Chern number to determine the topological properties on the  $k_x$-$k_y$ plane. For a given $k_z$, we can compute the Chern number $n_c(k_z) = -\frac{1}{2\pi}\iint dk_xdk_y\Omega_+^z$ along the $z$ direction. We find for $-k_W<k_z<0$ or $0<k_z<k_W$, the Chern number $n_c=1$; while for other case but $k_z\ne 0$, $n_c(k_z) = 0$. Such a non-zero Chern number corresponds to a $k_z$-dependent edge state. 

Now we apply an external magnetic field in the $z$-direction $\bm B=B\hat z$ to the coexisting topological system. The momentum plane perpendicular to the magnetic field will be quantized, while the momentum along the direction of the field is still a good quantum number. In the presence of a magnetic field, the new Hamiltonian could be obtained via the Peierls substitution $\bm k\rightarrow\bm k-e\bm A/\hbar$ with $e$ being the charge of an electron. If we choose the Landau gauge $\bm A=-By\hat x$, the new Hamiltonian of the coexisting system is shown as the following form:
\begin{align}
	H=&\begin{bmatrix}
		M_0-M_1(\hat{k}_x^2+\hat{k}_y^2+k_z^2)&Dk_z(\hat{k}_x-i\hat{k}_y)\\Dk_z(\hat{k}_x+i\hat{k}_y)&-M_0+M_1(\hat{k}_x^2+\hat{k}_y^2+k_z^2)
	\end{bmatrix},
\end{align}
with $\hat{k}_x=k_x-{y}/{\ell_B^2}$, $\hat{k}_y=-i\partial_y$, and $\ell_B=\sqrt{\hbar/|eB|}$ being the magnetic length. We can introduce two ladder operators $a^\dag=({\ell_B}/{\sqrt{2}})(k_x-{y}/{\ell_B^2}+\partial_y)$ and $a=({\ell_B}/{\sqrt{2}})(k_x-{y}/{\ell_B^2}-\partial_y)$, which satisfy $[a,a^\dag]=1$.  This  Hamiltonian is then rewritten as
\begin{align}
	H=&\begin{bmatrix}
		-M_{k_z}-\omega_c\left(a^\dag a+\frac{1}{2}\right)&\omega_za\\\omega_za^\dag&M_{k_z}+\omega_c\left(a^\dag a+\frac{1}{2}\right)
	\end{bmatrix},
\end{align}
with the frequencies $\omega_c={2M_1}/{\ell_B^2}$ and $\omega_z=\sqrt{2}Dk_z/{\ell_B}$, and $M_{k_z}=-M_0+M_1k_z^2$. Therefore the Landau energies and wavefunctions of the Hamiltonian $H$ for the index $\nu\ge 1$ are
\begin{align}
    E_{\lambda\nu}=\frac{1}{2}\omega_c+\lambda\sqrt{\left[M_0-M_1k_z^2-\nu\omega_c\right]^2+\nu\omega_z^2},
\end{align}
\begin{align}
	\psi_{+\nu}=\frac{e^{i(k_xx+k_zz)}}{\sqrt{L_xL_z}}\begin{bmatrix}
		\sin\frac{\alpha_\nu}{2}\phi_{\nu-1} \\
		\cos\frac{\alpha_\nu}{2}\phi_{\nu} \\
	\end{bmatrix},\\
     \psi_{-\nu}=\frac{e^{i(k_xx+k_zz)}}{\sqrt{L_xL_z}}\begin{bmatrix}
		\cos\frac{\alpha_\nu}{2}\phi_{\nu-1} \\
		-\sin\frac{\alpha_\nu}{2}\phi_{\nu} \\
	\end{bmatrix},
\end{align}
while for $\nu=0$,
\begin{align}
E_0=-M_0+M_1k_z^2+\frac{1}{2}\omega_c,
\end{align}
\begin{align}
	\psi_{0}=\frac{e^{i(k_xx+k_zz)}}{\sqrt{L_xL_z}}\begin{bmatrix}
		0 \\
		\phi_0 \\
	\end{bmatrix}.
\end{align}
Here $L_xL_z$ is the area of the sample and $\tan\alpha_\nu={\sqrt{\nu}\omega_z}/({\nu\omega_c+M_{k_z}})$. $\phi_\nu$ is the usual harmonic oscillator eigenstate at the center $y_0=k_x\ell_B^2$ relating to the Hermite polynomials $\mathcal H_\nu(x)$
\begin{align}
    \phi_{\nu}(k_{x},y)=\frac{1}{\sqrt{\sqrt{\pi}2^\nu\nu !  \ell_{B}}} e^{-\left[\left(y-y_{0}\right)^{2} / 2 \ell_{B}^{2}\right]} \mathcal{H}_{\nu}\left(\frac{y-y_{0}}{\ell_{B}}\right).
\end{align}

\begin{figure}[tbp]
\centering
\includegraphics[width=1\columnwidth]{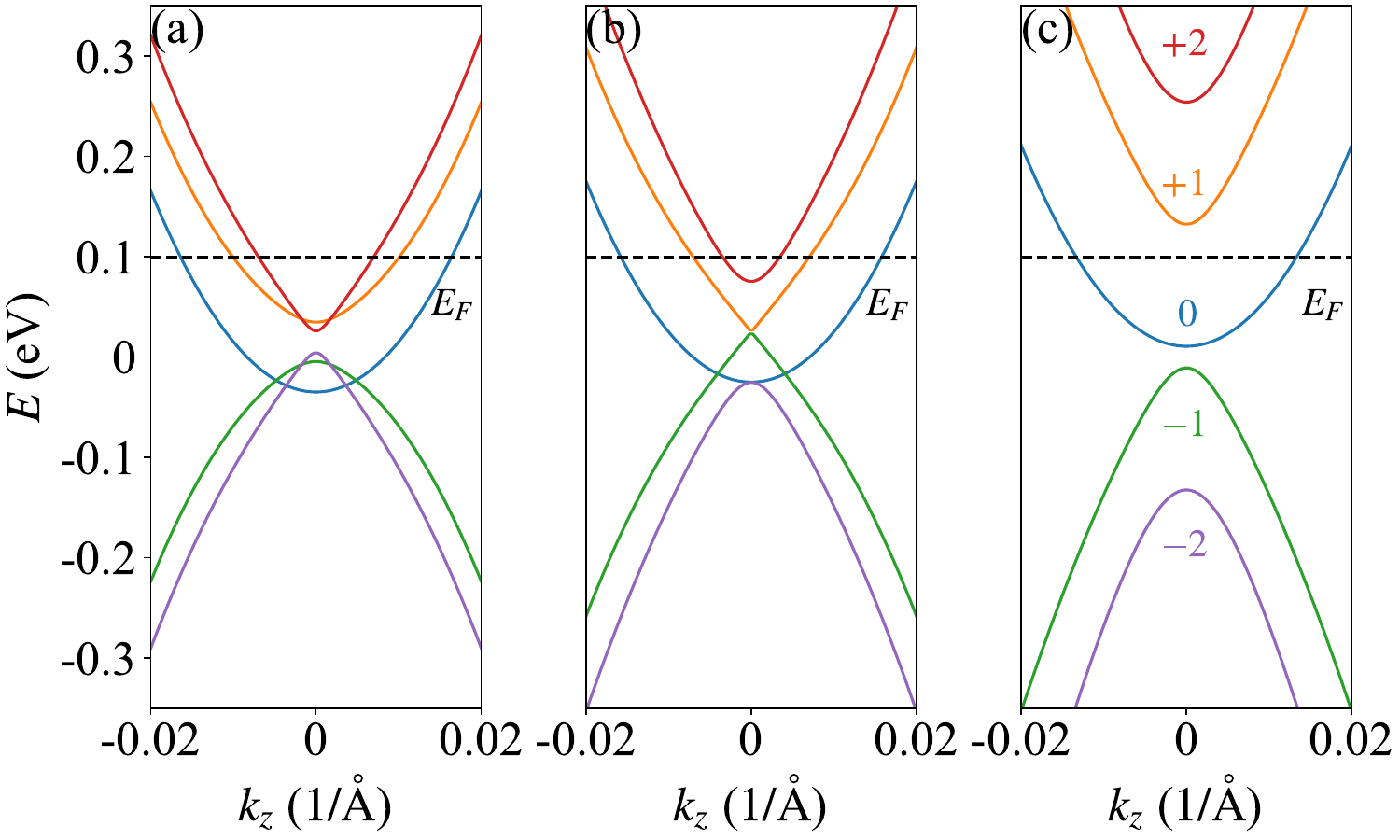}
\caption{The Landau energy bands of the system with indices $\nu=0, 1, 2$ for three different magnetic fields $B=2, 3.3, 8$ T. Here $M_0=0.05$ eV, $M_1=5$ eV nm$^2$, and $D=2M_1$.}
\label{fig-LL}
\end{figure}

In contrast to the two-node Weyl semimetal model, there is an additional frequency $\omega_z$ in the Landau bands. It relates to the parameter $D$ and is $k_z$-dependent, which is a key feature of the coexisting system. For the indices $\nu\ge 1$, two Landau energy levels $E_{\pm1,\nu}$ intersect at $k_z=0$ when the magnetic field $B_\nu=\hbar M_0/(2\nu |e| M_1)$. Figure \ref{fig-LL}(b) indicates the Landau bands at $B=B_1\simeq 3.3$ T, where the $+1$ band and the $-1$ band come into contact at $k_z=0$. This intersection is a feature of Landau levels in nodal-line semimetals in the presence of the magnetic field perpendicular to the nodal-line plane \cite{Li18prl}. But the zeroth Landau energy band $E_0$ is analogous to the two-node Weyl semimetal model \cite{Lu15Weyl-shortrange}, which is quadratic along the $k_z$ direction. 

Due to the $k_z$-dependent frequency $\omega_z$, the Landau bands exhibit a complexity in vivid contrast to the two-node Weyl model, especially at low magnetic fields. In the two-node Weyl model, Landau bands are distinctly separated and organized in the order of band indices \cite{Lu15Weyl-shortrange}. However, in this coexisting topological system, the Landau bands overlap, with even the lowest Landau band extending into the negative band (valence band) at low magnetic fields, as illustrated in Fig. \ref{fig-LL}(a) and (b). This band-crossing behavior is anticipated to give rise to unique magnetic transport phenomena, such as novel magnetoresistance oscillations \cite{WangCM2020prbrc}. Currently, our focus remains on the quantum limit, where only one Landau band is occupied. Fortunately, when the magnetic field is sufficiently large such that the bottom of the zeroth Landau band is higher than the top of the $-1$ Landau band, as depicted in Fig. \ref{fig-LL}(c), the bands become well-separated and ordered according to the Landau indices. This condition occurs when the magnetic field $B>\hbar M_0/(| e| M_1)$, twice the value of $B_1$. In such cases, if the Fermi energy $E_F$ intersects only the zeroth band, the coexisting topological system enters the quantum limit. Our subsequent calculations will focus on this extreme regime.

\section{longitudinal configuration}\label{longitudinal}
At first, we consider the longitudinal configuration, where the magnetic field, the electric field and the temperature gradient are all along the $z$-direction, $\bm B\parallel\bm E\parallel \nabla T\parallel\hat z$. The Mott relation is believed to be valid at temperatures close to 0 K \cite{Fu2022prb}. Hence, in this configuration we only need to calculate the longitudinal conductivity $\sigma_{zz}$. The resistivity $\rho_{zz}$ is just the reciprocal of the conductivity, $\rho_{zz}=1/\sigma_{zz}$. And the thermoelectric Seebeck coefficient $S_{zz}$ could be obtained from the coductivity via the Mott relation \cite{Fu2022prb,Ferreiros17prb,Jonson1980prb,Lundgren14prb}
\begin{align}
	S_{zz}=&\frac{\pi^2k_B^2T}{3e}\frac{1}{\sigma_{zz}}\frac{\partial\sigma_{zz}} {\partial E_F}.
\end{align}
Here $k_B$ is the Boltzmann constant, $T$ is the temperature and $E_F$ is the Fermi energy.

The longitudinal conductivity $\sigma_{zz}$ can be expressed as \cite{Lu15Weyl-shortrange,ZhangSB16njp,Fu2022prb,Li2023prb}
\begin{align}
	\sigma_{zz}=\frac{\hbar e^2}{2\pi V}\int_{-\infty}^\infty d\varepsilon\left[-\frac{\partial n_\text{F}(\varepsilon)}{\partial \varepsilon}\right]\text{Tr}\left[\hat v_z \hat G^A(\varepsilon)\hat v_z \hat G^R(\varepsilon)\right].
\end{align}
Here $V$ is the volume, $n_\text{F}(\varepsilon)$ is the Fermi-Dirac distribution function, $\hat G^{R/A}(\varepsilon)$ is the retarded (advanced) Green's function and $\hat v_z$ is the $z$-component of the velocity operator with the form
\begin{align}
	\hat v_z=\frac{1}{\hbar}\frac{\partial H}{\partial k_z}=\frac{1}{\hbar}\begin{bmatrix}
	-2M_1k_z&\frac{\sqrt{2}}{\ell_{B}}Da\\
	\frac{\sqrt{2}}{\ell_{B}}Da^\dag&2M_1k_z	
\end{bmatrix}.
\end{align}
Different from the two-node Weyl model, the off-diagonal element of the velocity operator $\hat{v}_z$ is non-zero due to the $Dk_z$ term in the Hamiltonian. However, for any band with non-zero index $\nu$, $\left \langle \psi_0 \right | \hat{v}_z \left | \psi_{\pm\nu} \right \rangle=0$. Thus in the quantum limit, where the Fermi energy only cuts the lowest Landau band $E_0$, only the zeroth band $E_0$ contributes to the conductivity in the longitudinal configuration. In this case only the velocity element $v_{k_z}=\left \langle \psi_0 \right | \hat{v}_z \left | \psi_0\right \rangle=2M_1k_z/\hbar$ involves in the expression leading to the longitudinal conductivity of this coexistence model analogous to the two-node Weyl model \cite{Lu15Weyl-shortrange,Li2023prb}:
\begin{align}
	\sigma_{zz}
	=2\frac{ e^2}{h }N_Lv_{k_F} \tau_\text{tr}^\text{0},	
\end{align}
where the Landau degeneracy $N_L=1/2\pi\ell_B^2$, the Fermi group velocity $v_{k_F}=2M_1k_F/\hbar$, and the Fermi wave vector $k_F$ is the positive intersection of the Fermi energy and the $0$th Landau band. The transport time at the Fermi energy, denoted as $\tau^{0}_\text{tr}$, includes an additional factor $(1-v_{ k_{z}^{\prime}}^{z}/v_{k_F}^z)$ from the vertex correction in its expression, distinct from the particle lifetime \cite{Mahan1990, Fu2022prb}. This factor introduces a selection rule, allowing only $k_{z}^{\prime}=-k_F$ to contribute to the transport time. Finally, $\tau_\text{tr}^0$ has the form
\begin{align} 	
	\frac{\hbar}{\tau_\text{tr}^\text{0}}	=&\frac{n_{i}}{M_1k_F} \sum_{k_{x}^{\prime},\bm q}|u({\bm q})|^2 e^{-\frac{q_\|^2\ell_B^2}{2}}\delta_{q_x,k_x-k_x'}\delta_{q_z,2k_F}.
\end{align}
Here $n_{i}$ is the impurity density, $u({\bm q})$ represents the Fourier transform of the scattering potential, the wave vector $\bm q=(q_x,q_y,q_z)$, and $q_\|^2=q_x^2+q_y^2$.

We consider two popular types of elastic electron-impurity scatterings: the Gaussian potential and the screened Coulomb potential. The former characterizes scenarios where scattering intensity diminishes rapidly with distance from the scattering center, while the screened Coulomb potential represents situations where charged impurities induce substantial screening effects. The Fourier transform of the Gaussian potential is $u(\bm q)=u_0e^{-q^2d^2/2}$, where $u_0$ measures the scattering strength, and $d$ is a parameter that describes the range of the scattering potential. If we define $V_{G}=n_{i}u_0^2$, then the transport time is
\begin{align}
    \frac{\hbar}{\tau_\text{tr}^{0,G}}
    =\frac{V_G}{2\pi M_1k_F(2d^2+\ell_B^2)}e^{-4k_F^2d^2}.
\end{align}
Therefore, the conductivity has the form
\begin{align}
	\sigma_{zz}
     =\frac{ e^2}{h }\frac{(2M_1k_F)^2}{V_G}\left(1+\frac{2d^2}{\ell_B^2}\right)e^{4k_F^2d^2}.\label{conductivity_G}
\end{align}
Since only $k_F$ depends on the Fermi energy, the Seebeck coefficient $S_{zz}$ is given by
\begin{align}
		S_{zz}=\frac{\pi^2k_B^2T}{3e}\frac{1}{M_1k^2_F}\left(1+4d^2k_F^2\right).\label{S_G}
\end{align}
Further, the thermoelectric conductivity could be found via $\alpha_{zz}=\sigma_{zz}S_{zz}$. 

For the $\delta$-form short-range scatters $d=0$, $\sigma_{zz}\propto k_F^2$, and $S_{zz}\propto k_F^{-2}$, so the thermoelectric conductivity is a constant irrespective of the magnetic field
\begin{align}
	\alpha_{zz}=\frac{4\pi^2}{3}\frac{ek_B^2T}{h}\frac{M_1}{V_G}.\label{constant_al}
\end{align}
It does not rely on carrier density or Fermi energy, indicating that $\alpha_{zz}$ attains plateaus under sufficiently strong magnetic fields. Previously observed in Dirac systems, the Hall thermoelectric conductivity demonstrates plateau behavior in the extreme quantum limit \cite{Kozii2019prb}. In this study, we identify the plateau characteristics of longitudinal thermoelectric conductivity. This phenomenon is expected to be applicable to all Weyl-like semimetals. 




Now we move to the analysis of the screened Coulomb potential. The Fourier transform is $u(\bm q)=\frac{e^{2}}{\epsilon_0\epsilon_r (q^2+\kappa^2)}$ with $\epsilon_0$ and $\epsilon_r$ being the dielectric constants of the vacuum and the material. The $\kappa$ from the standard random phase approximation is $\kappa^{2}=\frac{e^{2}}{2\pi\epsilon_0\epsilon_r} \frac{1}{2 \pi \ell_{B}^{2}}\frac{1}{M_1k_F}$. In this case, the transport time is
\begin{align}
\frac{\hbar}{\tau_\text{tr}^{0,C}}
     =&\frac{V_{C}}{M_1k_F}\frac{\ell_B^2}{8\pi}F_1(c_1).
\end{align}
Here $V_{C}=n_{i}e^4/(\epsilon_0^2\epsilon_r^2)$, the function $F_1(x)=1/x-e^xE_1(x)$ with $E_1(x)=\int_{x}^\infty dt{e^{-t}}/{t}$ being an exponential integral function, and $c_1=(4k_F^2+\kappa^2)\ell_B^2/2$. Therefore, the conductivity has the form
\begin{align}
	\sigma_{zz}
     =&\frac{ e^2}{h }\frac{(4M_1k_F)^2}{V_C\ell_B^4}\frac{1}{F_1(c_1)},\label{eqn:sgzz}
\end{align}
meanwhile, the Seebeck coefficient can be expressed as 
\begin{align}
S_{zz}
  =&\frac{\pi^2k_B^2T}{3e}\frac{1}{M_1k_F^2}\left[1-\frac{k_F}{2}\frac{\partial \ln F_1(c_1) }{\partial k_F}\right].\label{eqn:szz}
\end{align}

The function $F_1(x)$ has the asymptotic behavior: $F_1(x)= 1/x$ when $x\ll1$, while $F_1(x)=1/ x^{2}$ when $x\gg1$. Hence, for $c_1\ll1$, $\sigma_{zz}\propto k_F^2(4k_F^2+\kappa^2)/\ell_B^2$, $S_{zz}\propto k_F^{-2}(16k_F^2+\kappa^2)/(4k_F^2+\kappa^2)$, and $\alpha_{zz}\propto {(16k_F^2+\kappa^2)}/{\ell_B^2}$; while for $c_1\gg1$, we find $\sigma_{zz}\propto k_F^2 (4k_F^2+\kappa^2)^2$, $S_{zz}\propto(4k_F^2+\kappa^2)^{-1}$, $\alpha_{zz} \propto k_F^2(4k_F^2+\kappa^2)$. The $\kappa$ relies on the Fermi wave vector $k_F$. Therefore, the magnetic field dependence of these three transport quantities strongly depends on how the Fermi wave vector relies on the field. It is known that $k_F$ is related to the filling state of electrons. When the Fermi energy is fixed, the Fermi wave vector is given by
\begin{align}
	k_F=\sqrt{\frac{1}{M_1}({E_F+M_0-{\omega_c}/{2}})}.
\end{align}
The $B$-dependence is in the frequency $\omega_c$. Another choice is to keep the carrier density occupying the lowest energy band unchanged, that is to say $N_e=\sum_{k_x,k_z}\Theta(E_F-E_{0})$ is a constant, which leads to
\begin{align}
	k_F=2\pi^2\ell_B^2N_e.
\end{align}
Now the $k_F$ is inversely proportional to the magnetic field.

\begin{figure}[tbp]
\centering
\includegraphics[width=1\columnwidth]{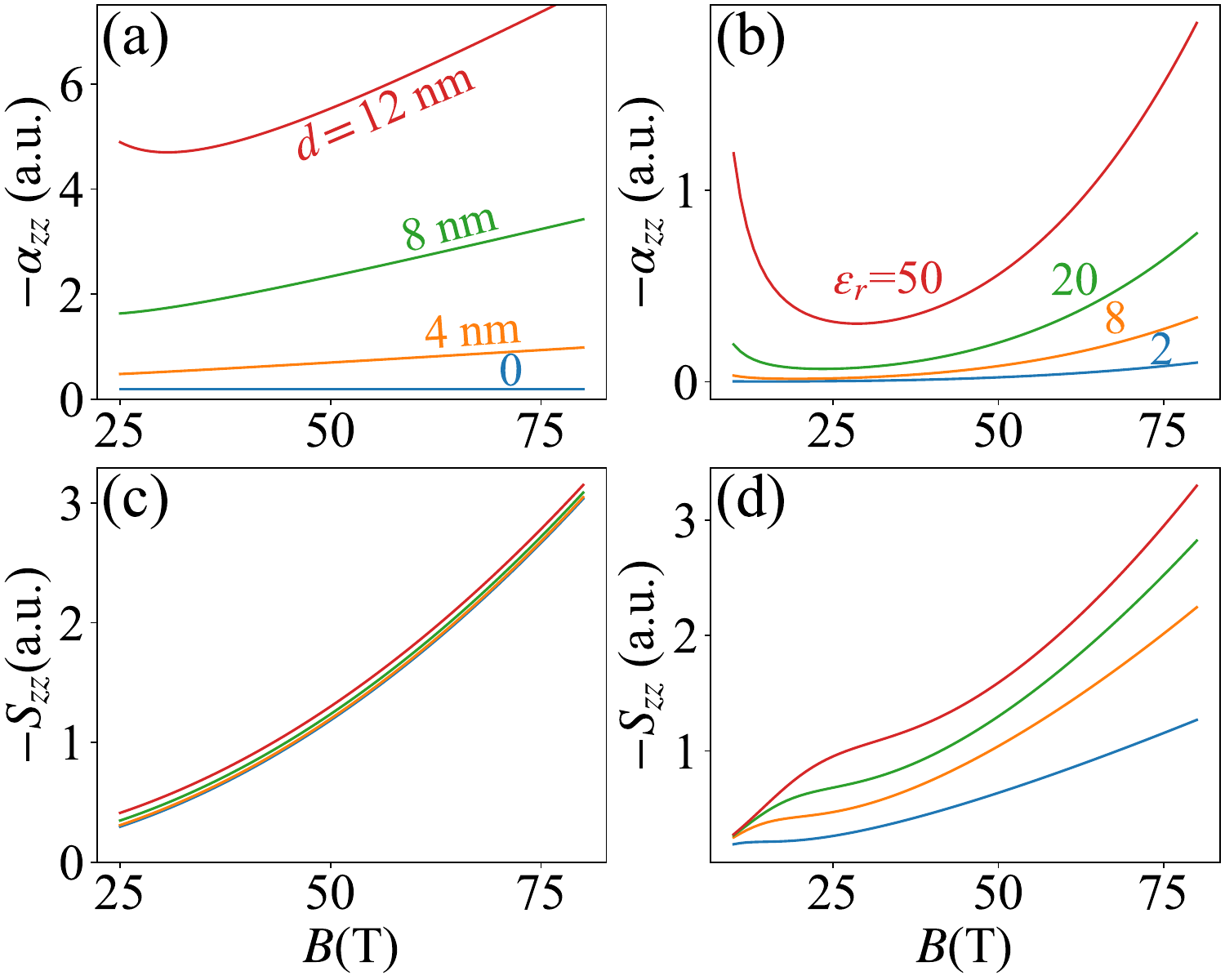}
\caption{The thermoelectric conductivity $\alpha_{zz}$ and the Seebeck coefficient of the coexisting system are plotted as functions of the magnetic field $\bm B$ at Gaussian [(a) and (c)] and Coulomb [(b) and (d)] potentials for fixed carrier density $N_e=5\times10^{22}$ m$^{-3}$. The four curves in (a) and (c) are taken for $d=0,4,8,12$ nm, and  four curves in (b) and (d) correspond to $\epsilon_r=2,8,20,50$. Here $V_\text{G}=10^{-4}$ eV$^2$nm$^3$, and $V_\text{C}=0.1$ eV$^2/$nm.} \label{fig-Longitudinal}
\end{figure}

Since the conductivity in the longitudinal configuration is the same as the one of the two-node Weyl model, which has been carefully studied before \cite{Li2023prb,ZhangSB16njp}, here we focus on the thermoelectric quantities of this coexisting model. Figure \ref{fig-Longitudinal} shows the longitudinal thermoelectric transport contributions of the coexistence topological semimetal as a function of magnetic field at a fixed carrier density for the Gaussian-type [(a) and (c)] and Coulomb-type [(b) and (d)] elastic scatterings. For $\delta$-form scattering ($d=0$ in the Gaussian potential), the thermoelectric conductivity $\alpha_{zz}$ remains constant independent of the magnetic field, aligning with Eq. \eqref{constant_al}. However, for Gaussian-type scattering with nonzero distance, it linearly depends on the magnetic field at high field strengths. This behavior can be understood from Eqs. \eqref{conductivity_G} and \eqref{S_G}: for short distances, such as $dk_F\ll1$, the thermoelectric conductivity exhibits a weak linear relationship with $B$, where $\alpha_{zz}=\sigma_{zz}S_{zz}\propto1+2d^2/\ell_B^2$. For long distances (e.g., $d=12$ nm), $\alpha_{zz}$ may first decrease and then increase with the increment of the field. At short distances where $dk_F\ll1$, several curves ($d=2,4,8$ nm) of the Seebeck coefficient almost exactly coincide and grow parabolically with the magnetic field since, in this case, $S_{zz}\simeq\frac{\pi^2k_B^2T}{3e}\frac{1}{M_1k^2_F}\propto B^2$. As the distance increases, deviations begin to appear.

\begin{figure}[tbp]
	\centering
	\includegraphics[width=1\columnwidth]{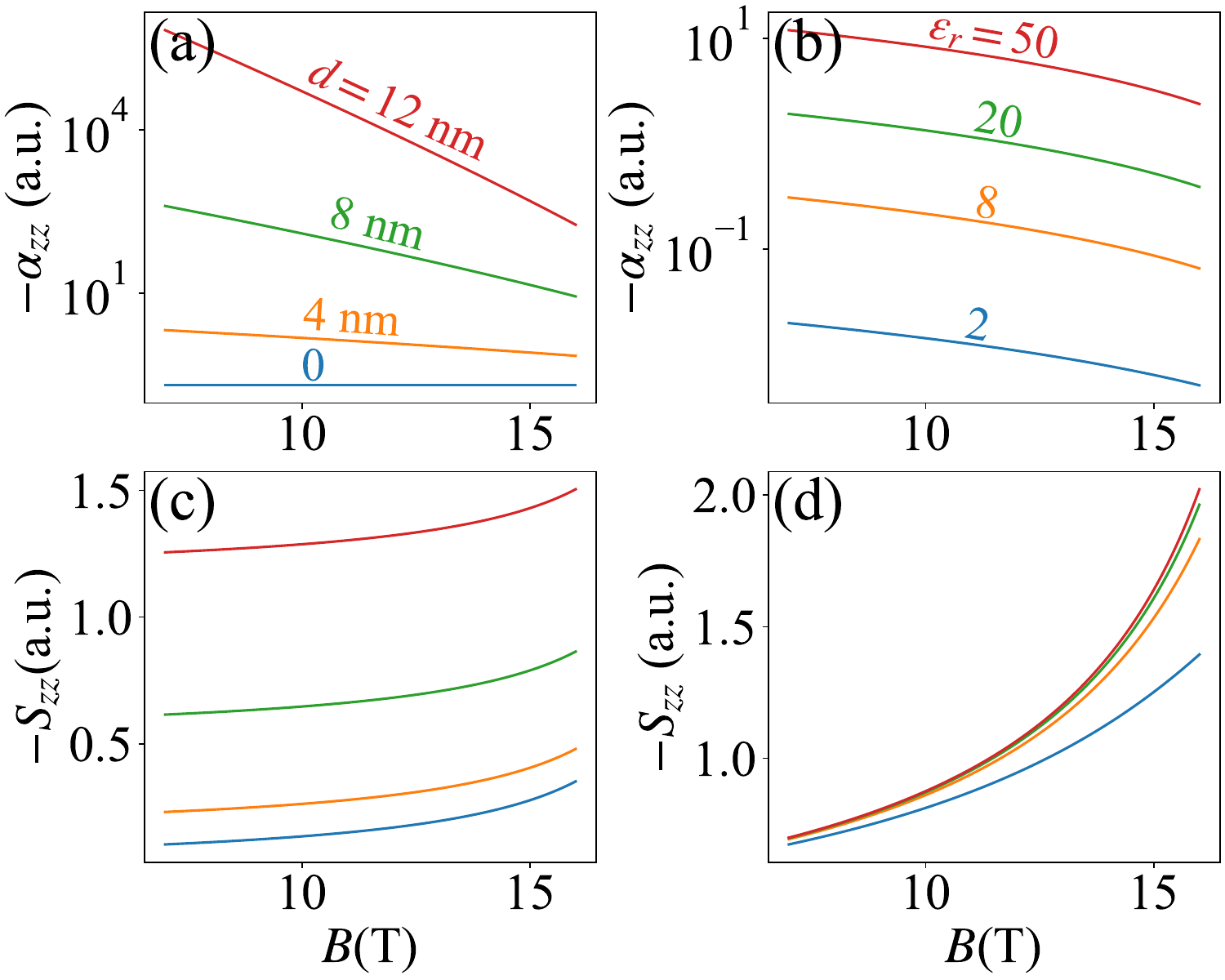}
	\caption{The thermoelectric coefficients $\alpha_{zz}$ and the Seebeck coefficient of the coexisting system versus the magnetic field at Gaussian [(a) and (c)] and screened Coulomb [(b) and (d)] potentials at fixed Fermi energy $E_F=2M_0$. The other parameters are the same as Fig. \ref{fig-Longitudinal}.} \label{fig-LongitudinalE}
\end{figure}

For the screened Coulomb scattering, we first discuss the results with large relative dielectric constant ($\epsilon_r =20,50$). The thermoelectric coefficient $\alpha_{zz}$ first decreases and then increases with the field shown in Fig. \ref{fig-Longitudinal}(b). For this case, the small field limit corresponds to $4k_F^2\gg\kappa^2$, but the large field one corresponds to $4k_F^2\ll\kappa^2$. We find that: (i) in the small magnetic field limit, the thermoelectric coefficient $\alpha_{zz}$ transitions form $\propto B^{-4}$ at $2k_F^2\ell_B^2\gg1$ to $\propto B^{-1}$ at $2k_F^2\ell_B^2\ll 1$, but $S_{zz}$ is proportional to $B^2$ with the increment of the field according to the previous analysis. (ii) For the strong magnetic field limit, Eq. \eqref{eqn:szz} can be simplified to $S_{zz}\simeq(2M_1k_F^2)^{-1}(\ell_B^2\kappa^2/2+1)^{-1}$ with the help of the approximation $F(c_1)\simeq 1/[c_1(c_1+1)]$. 
Hence, we find that $S_{zz}$ transitions from $\propto B^2$ at $\ell_B^2\kappa^2/2\ll 1$ to $\propto B$ at $\ell_B^2\kappa^2/2\gg1$ with the growth of the field indicating in Fig. \ref{fig-Longitudinal}(d). However, the thermoelectric coefficient $\alpha_{zz}= \sigma_{zz}S_{zz}$ is always proportional to $B^3$.
Now we move to the low relative dielectric constant case ($\epsilon_r =2,8$). In this case, the strong screening condition ($4k_F^2\ll\kappa^2$) is satisfied in the whole field regime.
The smaller dielectric constants allow us to witness the linear growth of $S_{zz}$ within smaller magnetic fields, as indicated by the orange and blue curves in Fig. \ref{fig-Longitudinal}(d). Similarly, $\alpha_{zz}$ maintains its cubic relationship with the magnetic field.
Hence, the Gaussian or Coulomb scattering with a large dielectric constant is intentional for obtaining a large Seebeck coefficient.

\begin{figure*}[tbp]
	\centering
	\includegraphics[width=\textwidth]{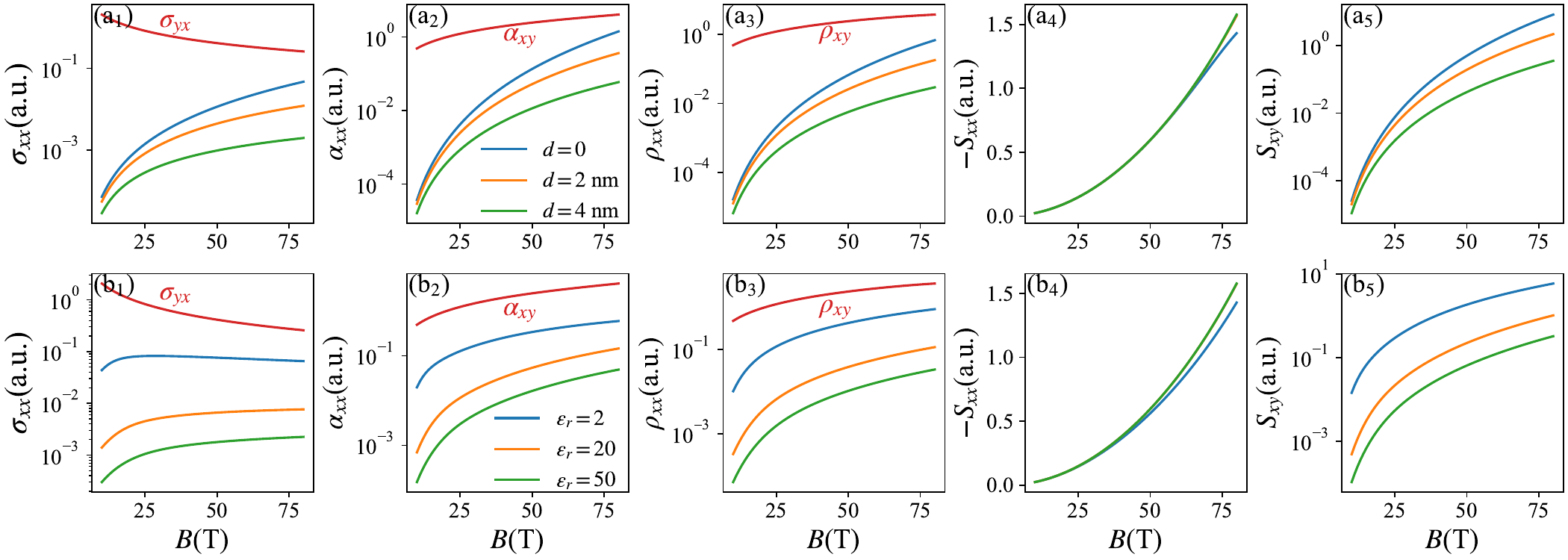}
	\caption{The conductivities, thermoelectric conductivity $\alpha_{xx}$, resistivities, Seebeck coefficient $S_{xx}$ and Nernst coefficient $S_{xy}$ of the coexistence topological semimetals at fixed carrier density $N_e=5\times10^{22}$ m$^{-3}$ as functions of magnetic field under Gaussian ($a_1,a_2,a_3,a_4,a_5$) and screened Coulomb ($b_1,b_2,b_3,b_4,b_5$) scatterings in the transverse case. The red curves in ($a_1$), ($a_2$), and ($a_3$) [and ($b_1$), ($b_2$), and ($b_3$)] represent $\sigma_{yx}$, $\alpha_{xy}$, and $\rho_{xy}$. Here $V_\text{G}=10^{-4}$ eV$^2$nm$^3$, $V_\text{C}=0.1$ eV$^2/$nm, and $T=0.5$ K. The other model parameters are the same as Fig. \ref{fig-LL}.}\label{fig-Transverse}
\end{figure*}

At temperatures near 0 K, the chemical potential $\mu \approx E_F$. Alternatively, in experiments where a fixed Fermi energy is often considered, we select the Fermi energy $E_F=2M_0$ to position the system in the quantum limit. Figure \ref{fig-LongitudinalE} depicts the variations of thermoelectric conductivity and Seebeck coefficients with magnetic field for two scattering potentials at a fixed Fermi energy. In this scenario, the magnetic field cannot be excessively applied, as a too-large magnetic field would elevate the lowest energy band $E_0$ beyond the Fermi energy, contrary to the fixed carrier density case.
The results for Gaussian-type scattering are presented in (a) and (c). For $\delta$-form scattering ($d=0$), the thermoelectric coefficient $\alpha_{zz}$ remains constant. When the distance is nonzero ($d\neq0$), $\alpha_{zz}$ increases rapidly with $d$ and decreases swiftly with an increasing magnetic field due to the exponential relation $\alpha_{zz}\propto e^{4k_F^2d^2}$. The Seebeck response $S_{zz}$ increases with the field following $S_{zz}\propto {k_F^{-2}}$ for $\delta$-form scattering. The curves of $S_{zz}$ with $d\neq0$ are essentially shifted by a field-independent constant $\frac{\pi^2k_B^2T}{3e}\frac{4d^2}{M_1}$ from the curve with $d=0$, in accordance with Eq. \eqref{S_G}.
For screened Coulomb scattering, all thermoelectric conductivities with different $\epsilon_r$ decrease with the magnetic field. This decrement of $\alpha_{zz}$ for the fixed Fermi energy contrasts with the case of fixed carrier density. Regarding the Seebeck coefficient, where the dependence on $\epsilon_r$ only arises from $\kappa$, its value grows rapidly with the field, as shown in Fig. \ref{fig-LongitudinalE}(d).

\section{transverse configuration}\label{transverse}
In the transverse configuration, the magnetic field remains aligned along the $z$-direction, maintaining the validity of the previously derived Landau energy bands and eigenstates. However, the electric field (temperature gradient) is now chosen to be along the $x$ axis perpendicular to the magnetic field. In this way the conductivities include the longitudinal one $\sigma_{xx}$ and the Hall one $\sigma_{xy}$. For the longitudinal conductivity $\sigma_{xx}$, we could still use the Kubo formula. But the velocity should be taken as the component in the $x$ direction. Since the $z$-direction magnetic field quantizes the energy in the $x$-$y$ plane, the needed expectation value in the quantum limit $v_x=\langle \psi_{0}|\hat v_{x}|\psi_{0}\rangle$ is zero. Hence, the associated velocity elements come from the off-diagonal ones. The nonzero conductivity $\sigma_{xx}$ originates from the interband velocity $v_{0,\pm1}=\langle \psi_{0}|\hat v_{x}|\psi_{\pm1}\rangle$ \cite{Lu15Weyl-shortrange}. $v_{0,+1}$ is expressed as
\begin{align}
v_{0,+1}&= \frac{1}{\hbar}\left(Dk_z\sin\frac{\alpha_1}{2}+\frac{\sqrt{2}}{\ell_B}M_1\cos\frac{\alpha_1}{2}\right).
\end{align}
$v_{0,-1}$ can be obtained by exchanging sine and cosine, as well as changing the sign from positive to negative in $v_{0,+1}$. The expression for the longitudinal conductivity in the quantum limit, $\sigma_{xx}$, is written as:
\begin{align} 	
	\sigma_{xx} =&\frac{\hbar e^2}{2\pi L_z}\frac{1}{2\pi\ell_B^2}\sum_{m = \pm1,k_z}v_{0,m}G^A_{m}(E_F)v_{m,0}  G^R_{0}(E_F),\label{sigmaxx} 
\end{align}
where the retarded and advanced Green's functions at the Fermi energy are
\begin{align} 
	G^R_{0}(E_F)=&\frac{1}{E_F-E_{0}+i\frac{\hbar}{2\tau_{0}}},\\ 
	G^A_{\pm1}(E_F)=&\frac{1}{E_F- E_{\pm 1}-i\frac{\hbar}{2\tau_{\pm 1}}}.
\end{align}
Here $\tau_0$ and $\tau_{\pm1}$ are the lifetimes, where $\tau_{\pm1}$ describes the virtual process going back and forth between bands $E_0$ and $E_{\pm1}$. We keep only the real part of $\sigma_{xx}$ since the fields considered here are all static. In the weak scattering limit $E_F-E_{m}\gg{\hbar}/{(2\tau_{m})}$, the conductivity $\sigma_{xx}$ becomes
\begin{align}
\sigma_{xx}\simeq&\frac{\hbar e^2}{\pi L_z}\frac{1}{2\ell_B^2}\sum_{m,k_z}\left[\frac{\hbar}{2\tau_{m}}\frac{(v_{0,m})^2}{\left(E_F-E_{m}\right)^2  }\delta\left(E_F-E_{0}\right)\right].
\end{align}
The lifetime could be calculated in the Born approximation \cite{Fu2022prb}, where both the $k_z=k_F$ and $k_z=-k_F$ terms contribute in contrast to the transport time. Hence, they are given by
\begin{align}
\frac{\hbar}{\tau_{\pm1}}
=&\frac{n_{i}(1\pm\cos\alpha_{1F})}{4M_1k_F}\sum_{k_{x}^{\prime}}\sum_{q_x,q_y}\big[|u({q_x,q_y,0})|^2\nonumber\\&+|u({q_x,q_y,2k_F})|^2\big] e^{-\zeta}\zeta \delta_{q_x,k_x-k_x'},
\end{align}
with $\zeta={q_\|^2\ell_B^2}/{2}$. $\alpha_{1F}$ is the value of $\alpha_1$ at $k_z=k_F$. The behaviour of $\tau_{\pm1}$ exhibits variance depending on the scattering potentials, leading to distinct transport quantities.

For the Gaussian potential, the relaxation time $\tau_{\pm1}^G$ is given by a specific expression
\begin{align}
\frac{\hbar}{\tau_{\pm1}^G}
=&\frac{V_G\ell_B^2(1\pm\cos\alpha_{1F})}{8\pi M_1k_F(2d^2+\ell_B^2)^2}\left(1+e^{-4k_F^2d^2}\right).
\end{align}
Therefore, the corresponding $\sigma_{xx}$ is of the form
\begin{align}
\sigma_{xx}^G
=&\frac{e^2}{h}\frac{V_G\ell_B^2(1+\cos^2\alpha_{1F})}{64\pi^2M_1^2k_F^2(2d^2+\ell_B^2)^2}\left(1+e^{-4k_F^2d^2}\right).
\end{align}
The thermoelectric conductivity $\alpha_{xx}$ can be determined through the Mott relation $\alpha_{xx}=\frac{\pi^2k_B^2T}{3e}\frac{\partial\sigma_{xx}}{\partial E_F}$. It's worth noting that the material parameter $D$ affects conductivity via the factor $1+\cos^2\alpha_{1F}$. This factor only fluctuates between 1 and 2, hence, its impact on the overall magnetic field dependence of conductivity is relatively minor. Firstly, we disregard the influence of this factor. Additionally, when examining the magnetic field-dependent analytical behavior of transport quantities, we specifically consider the scenario of short-distance impurities, where $e^{-4d^2k_F^2}\approx 1$. In the fixed carrier density situation, we find $\sigma_{xx}^G\propto B$ and $\alpha_{xx}^G\propto B^3$ for short magnetic length (large magnetic field) $2d^2\gg\ell_{B}^2$, but $\sigma_{xx}^G\propto B^3$ and $\alpha_{xx}^G\propto B^5$ for long magnetic length (small magnetic field) $2d^2\ll\ell_{B}^2$.

\begin{figure}[tb]
	\centering
	\includegraphics[width=\columnwidth]{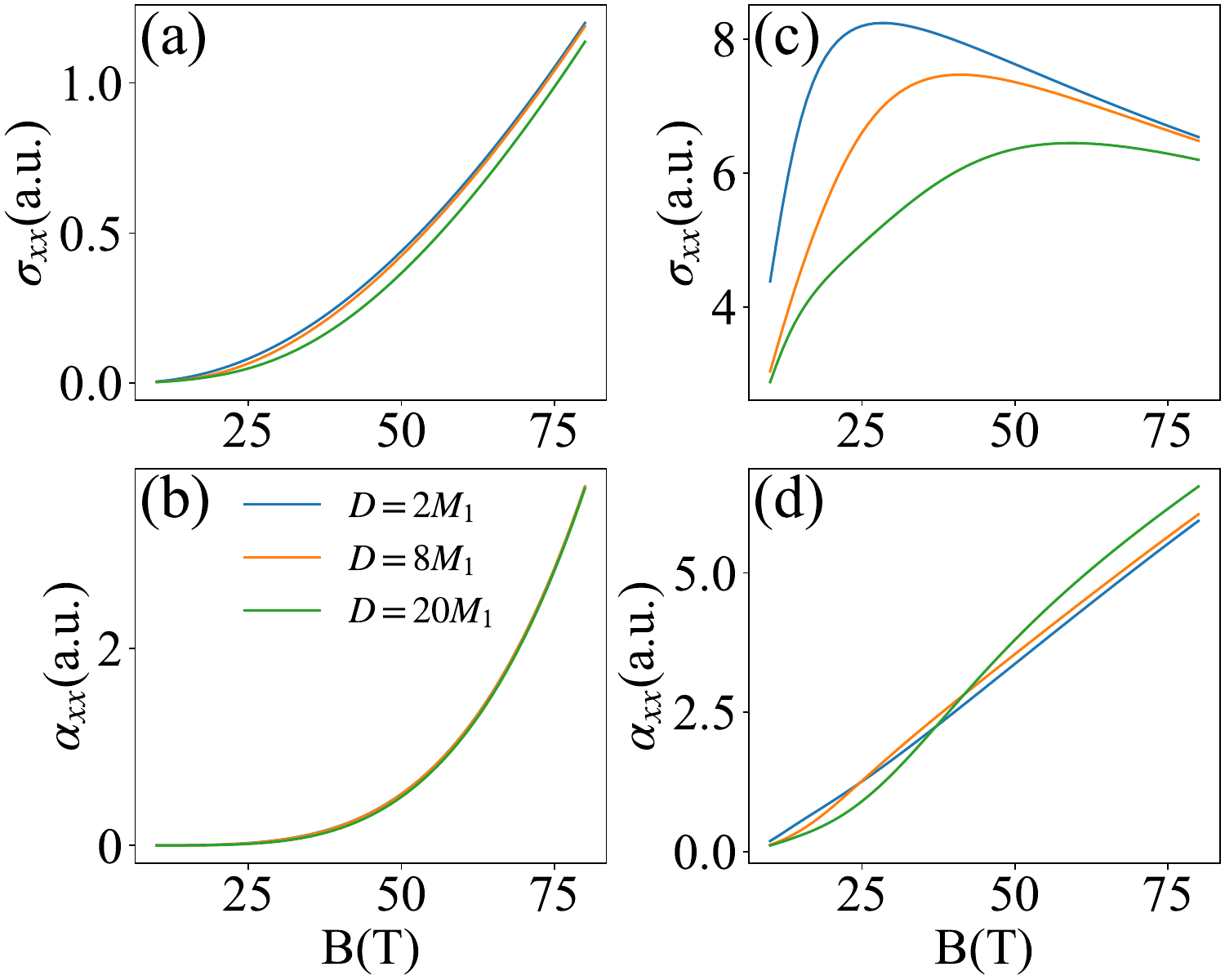} 
	\caption{The conductivities and thermoelectric conductivity $\alpha_{xx}$ of the transverse case as functions of magnetic field at fixed electron density $N_e=5\times10^{22}$ m$^{-3}$ for different $D=2M_1,8M_1,20M_1$ under Gaussian [(a) and (b)] and screened Coulomb [(c) and (d)] scatterings. Here $d=2$ nm and $\epsilon_r=2$. The other parameters are the same as Fig. \ref{fig-Transverse}.}\label{fig-Transverse-D}
\end{figure}

For the screened Coulomb potential, the lifetime becomes
\begin{align}
\frac{\hbar}{\tau_{\pm1}^{C}}
=&\frac{V_C{\ell_B}^2(1\pm\cos\alpha_{1F})}{16\pi M_1k_F}[F_2(c_1)+F_2(c_2)],
\end{align}
with $c_2=\ell_B^2\kappa^2/2$. Hence, the conductivity $\sigma_{xx}^C$ is expressed as 
\begin{align}
\sigma_{xx}^C
=&\frac{e^2}{h}\frac{V_\text{C}\ell_B^2(1+\cos^2\alpha_{1F})}{128\pi M_1^2k_F^2}[F_2(c_1)+F_2(c_2)].\label{CSxx}
\end{align}
We could also obtain $\alpha_{xx}$ from the Mott relation. Here the function $F_2(x)=-1+(1+x)e^xE_1(x)$.  $F_2(x)$ exhibits the asymptotic behavior $F_2(x)\propto x^{-2}$ at $x\gg1$, but $F_2(x)\approx \ln(1/x)-1-\gamma$ which is almost a constant independent of the magnetic field at $x\ll1$ with $\gamma$ being the Euler's constant. 
For this screened Coulomb potential, we also neglect the effect of the factor $1+\cos^2\alpha_{1F}$. We notice that: (i) if $c_2\gg 1$, the conductivity $\sigma_{xx}^C$ is inversely proportional to the magnetic field at both strong screening ($4k_F^2\ll\kappa^2$) and weak screening ($4k_F^2\gg\kappa^2$) for the fixed carrier density case. (ii) However, if $c_2\ll 1$, we have $\sigma_{xx}^C\propto B$ and $\alpha_{xx }^C\propto B ^3$.


For this transverse configuration ($\bm B\parallel \hat z$ and $\bm E\parallel \hat x$), another important quantity is the Hall conductivity $\sigma_{yx}$. For this three-dimensional system dispersing with $k_z$, it is known that each $k_z$ contributes a quantized Hall conductivity {$e^2/h$}, hence the total Hall conductivity is \cite{Lu15Weyl-shortrange}
\begin{align}
	\sigma_{yx}=&2\int_{0}^{k_F}\frac{dk_z}{2\pi}\frac{e^2}{h}=\frac{e^2}{h}\frac{k_F}{\pi}.\label{sigmaxy}
\end{align}
If we consider the case of the fixed carrier density $N_e$, we get the Hall conductance at zero temperature  $\sigma_{yx}=-{eN_e}/{B}$, this is the well-known classical Hall conductivity. Therefore, the resistivities $\rho_{xx}$ and $\rho_{xy}$ can be written as 
\begin{align}
    \rho_{xx}=&\frac{\sigma_{xx}}{\sigma_{xx}^2+\sigma_{yx}^2},\label{rhoxx}\\
    \rho_{xy}=&-\rho_{yx}=\frac{\sigma_{yx}}{\sigma_{xx}^2+\sigma_{yx}^2}.\label{rhoxy}
\end{align}

\begin{figure*}[tbp]
	\centering
	\includegraphics[width=\textwidth]{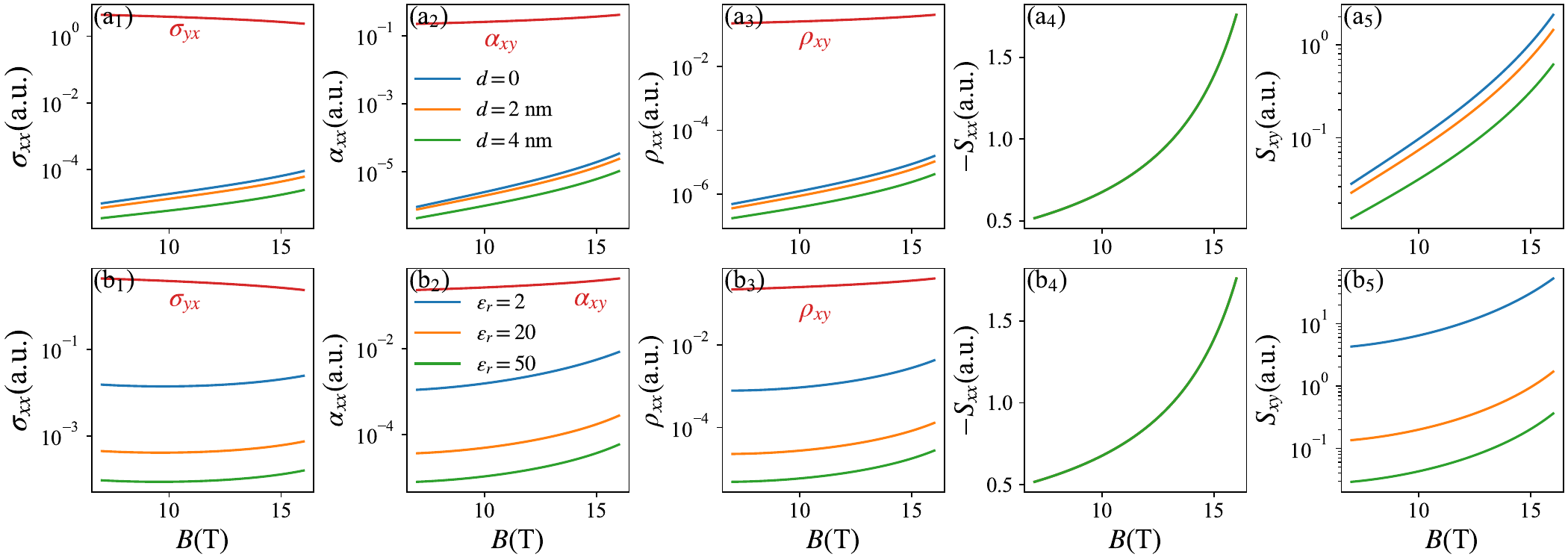}
	\caption{Same as Fig. \ref{fig-Transverse}, but at fixed Fermi energy $E_F=2M_0$.}
	\label{fig-Transverse-E}
\end{figure*}

In the presence of $\bm B\parallel \hat z$ and $-\nabla T\parallel \hat x$, due to the diffusion and drift motion of the carriers, the material generates voltage gradient, $E_x$ and $E_y$, respectively, along the $x$ and $y$ direction. Therefore, the Seebeck and Nernst coefficients can be expressed as
\begin{align}
    S_{xx}=&-E_x/|\nabla T|=\rho_{xx}\alpha_{xx}+\rho_{yx}\alpha_{xy},\label{Sxx}\\
    S_{xy}=&E_y/|\nabla T|=\rho_{xx}\alpha_{xy}-\rho_{yx}\alpha_{xx}\label{Sxy}.
\end{align}
Here $\alpha_{xy}$ is thermoelectric Hall coefficient, which in the quantum limit is given by \cite{Bergman2010prl,Kozii2019prb}
\begin{align}
\alpha_{xy}(B,T)&=-\frac{e}{2\pi\hbar L_z}\sum_{k_z}s\left(\frac{E_0(k_z)-E_F}{k_BT} \right),\label{alphaxy}
\end{align}
with the entropy per electron state being
\begin{align}
s(x)=k_B\left[\ln{(1+e^{x})}-\frac{x}{1+e^{-x}}\right].
\end{align}
If $M_1k_F^2=E_F-(-M_0+\omega_c/2)\gg k_BT$, that is to say, the distance between Fermi energy and the bottom of the zeroth Landau band is much greater than $k_BT$, $\alpha_{xy}$ could be calculated as
\begin{align} 	 	
	\alpha_{xy}\simeq&-\frac{ek_B^2T}{24\hbar M_1k_F }=\frac{e^2k_B^2TB}{48\pi^2\hbar^2 M_1N_e}.
\end{align}
It is proportional to the magnetic field when the carrier density is fixed, rather than a constant value, because the current lowest Landau level behaves like a Sch\"odinger particle \cite{Kozii2019prb}.

Figure \ref{fig-Transverse} shows the variation of $\sigma_{xx}$, $\alpha_{xx}$, $\rho_{xx}$, $S_{xx}$ and $S_{xy}$ with magnetic field at fixed carrier density for two scattering potentials [(a) Gaussian and (b) Coulomb potentials]. Additionally, employing Eqs. \eqref{sigmaxy} and \eqref{alphaxy} yields $\sigma_{yx}$ and $\alpha_{xy}$ displayed with red curves in the corresponding $\sigma_{xx}$ and $\alpha_{xx}$. For the $\delta$-form scattering (blue curve), the longitudinal conductivity $\sigma_{xx}$ experiences a faster increase than the cases with finite distances. This is due to $\sigma_{xx}\propto B^3$ under $\delta$ potential, while $\sigma_{xx}\propto B$ at large magnetic fields (short magnetic lengths) under Gaussian potential. Concerning the screened Coulomb potential, the conductivities increase with the magnetic field for both $\epsilon_r=20$ and $\epsilon_r=50$ in Fig. \ref{fig-Transverse}($b_1$), whereas the conductivity for $\epsilon_r=2$ (blue curve) initially increases and then decreases with the field. It's essential to note that logarithmic coordinates are used here, and a substantial decrease in conductivity is observable in normal coordinates [Fig. \ref{fig-Transverse-D}(c)]. The conductivity reduction is attributed to the fact that for small $\epsilon_r$, $c_1>c_2\gg1$ at large fields, causing the conductivity $\sigma_{xx}$ to tend towards $\propto B^{-1}$. In Figs. \ref{fig-Transverse}($a_2$) and ($b_2$), both longitudinal thermoelectric coefficients $\alpha_{xx}$ increase with the magnetic field. Contrary to the Dirac semimetal, the thermoelectric Hall conductivity $\alpha_{xy}$ exhibits an almost linear dependence on the field rather than forming a plateau, as seen in the red curves in Figs. \ref{fig-Transverse}($a_2$) and ($b_2$). This behavior arises because the zeroth band of this semimetal is similar to the Schr\"odinger particle \cite{Kozii2019prb}. The large positive magnetoresistance takes place in Figs. \ref{fig-Transverse}($a_3$) and ($b_3$) especially in the presence of Gaussian potential scattering. The Hall resistivity linearly increases with the field, nearly independent of the scattering. Consequently, we only plot one curve in Figs. \ref{fig-Transverse}($a_3$) and ($b_3$).

The Seebeck and Nernst coefficients for the two scattering potentials obtained through Eqs. \eqref{Sxx} and \eqref{Sxy} are shown in Figs. \ref{fig-Transverse}($a_4$), ($b_4$), ($a_5$), and ($b_5$), respectively. In this configuration, the rapid growth of the Seebeck and Nernst coefficients are observed for both scatterings. For the Gaussian potential with small distance ($d=2,4 $ nm) or the screened Coulomb potential with large dielectric constant ($\epsilon_r=20,50$), the Seebeck response can be approximated with the one in the dissipationless limit $S_{xx}\sim\alpha_{xy}/\sigma_{xy}\propto B^2$ independent of the scattering. This parabolic behavior arises from the linear dependence of $\alpha_{xy}$. In this scenario, $\sigma_{xx}\ll\sigma_{yx}$ at the given scattering parameters and electron density, making the first term in $S_{xx}$ negligible. Hence, the yellow and green curves in $S_{xx}$ almost completely overlap. The Seebeck response deviates from the dissipationless limit for the $\delta$-form scattering ($d=0$) or the screened Coulomb scattering with small relative dielectric constant ($\epsilon_r=2$) especially at larger magnetic field, where the $\rho_{xx}\alpha_{xx}$ in Eq. \eqref{Sxx} contributes. In contrast, the behavior of the Nernst coefficient strongly depends on the scattering, as two larger quantities, $\alpha_{xy}$ and $\rho_{yx}$, are in the two parts of $S_{xy}$. It grows rapidly with the magnetic field, and the small distance $d$ or the small dielectric constant $\epsilon_r$ could enhance its value.

   \begin{figure}[tbp]
	\centering
	\includegraphics[width=\columnwidth]{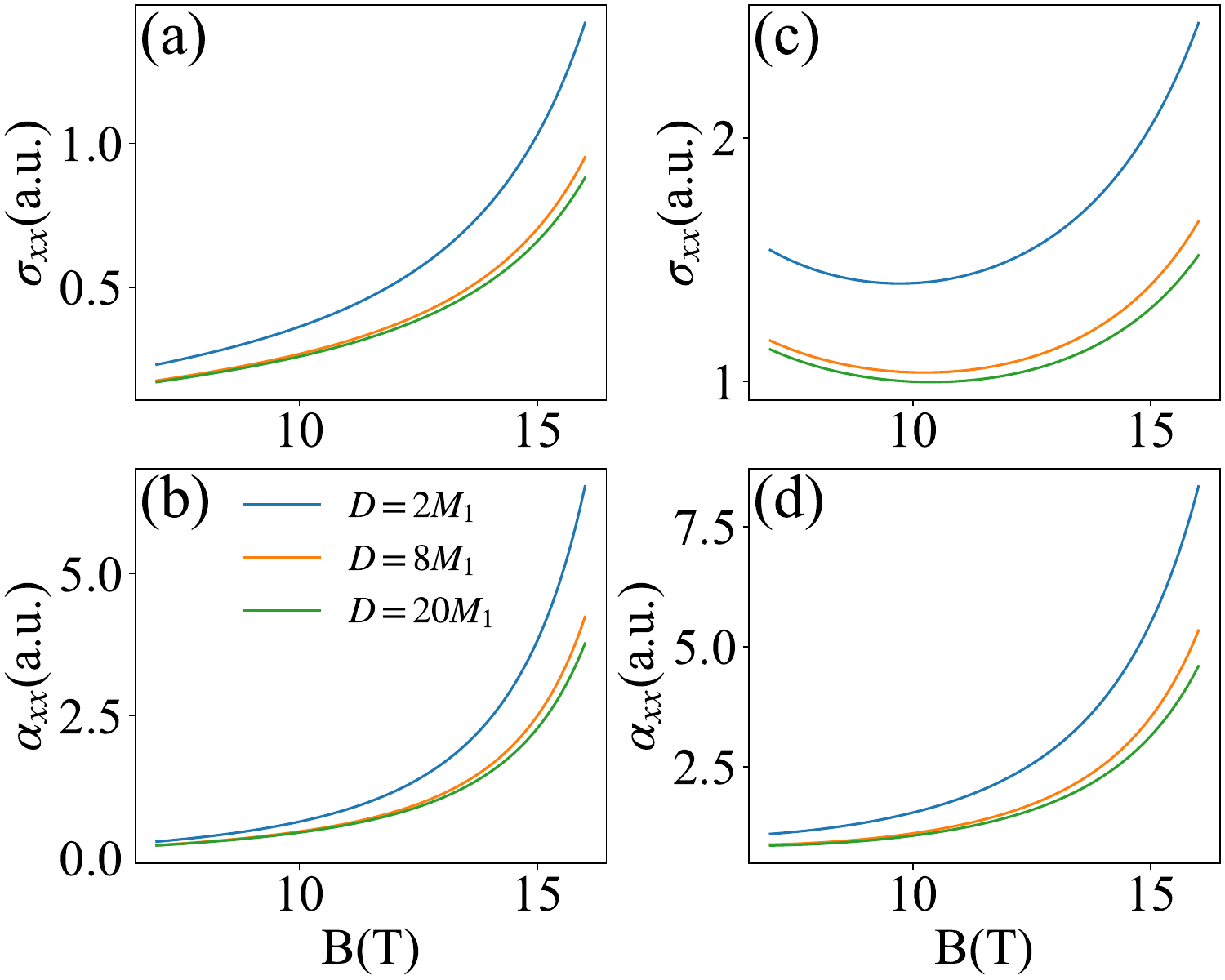} 
	\caption{Same as Fig. \ref{fig-Transverse-D}, but at fixed Fermi energy $E_F=2M_0$.}
	\label{fig-Transverse-DE}
\end{figure}

The model parameter $D$ distinguishes this coexistence model from the two-nodel Weyl model. In the transverse case, its effect will be revealed through the factor $\cos\alpha_{1F}$, which is illustrated in Fig. \ref{fig-Transverse-D}. We focus on the electric and thermoelectric conductivities for different scatterings [Gaussian scattering in (a) and (b), and screened Coulomb scattering in (c) and (d)]. It is observed that the effect of parameter $D$ is mainly concentrated near a moderate magnetic field since $\cos\alpha_{1F}$ nearly equals the constant 0 or 1 at small (large) magnetic fields. For the Gaussian potential, the parameter $D$ has a limited effect on the conductivity $\sigma_{xx}$, which slightly decreases with increasing $D$, and its magnetic field trend remains unchanged. However, the conductivity substantially decreases with increasing $D$ for the screened Coulomb potential, and the trend changes slightly at the same time. Overall, the influence of the parameter $D$ on thermoelectric conductivity $\alpha_{xx}$ is relatively small, as shown in Fig. \ref{fig-Transverse-D}(b) and (d).

The behaviors of these thermoelectric quantities at fixed Fermi energy are plotted in Fig. \ref{fig-Transverse-E} (different scattering potentials) and Fig. \ref{fig-Transverse-DE} (effect of the model parameter $D$). For the Gaussian potential, we observe that $\sigma_{xx}$ increases with the magnetic field, whereas it shows a slight decrease at fields less than 10 T for the screened Coulomb potential, in contrast to the case of fixed electron density. The longitudinal thermoelectric conductivity $\alpha_{xx}$ grows with the magnetic field for both scatterings, as shown in Figs. \ref{fig-Transverse-E}(a$_2$) and \ref{fig-Transverse-E}(b$_2$). In the entire range of the magnetic field, $\sigma_{xx}\ll\sigma_{yx}$ holds, allowing the Seebeck coefficient to be approximated with the one of the dissipationless limit irrespective of the scattering, proportional to the square of the magnetic field. However, the Nernst response strongly depends on the type of scatterings. From Fig. \ref{fig-Transverse-DE}, we can see that the model parameter $D$ could rapidly diminish the conductivity and thermoelectric conductivity, though it does not change the variation tendency. For the two cases of fixed Fermi level and fixed electron density, the overall trend of these transport quantities is similar, but there are significant differences in the specific magnetic field-dependent behavior.

\section{summary}\label{summary}
To conclude, we investigate the Landau bands of a coexisting topological system as well as discuss its thermoelectric transport properties in the quantum limit by using the linear response theory. Such a system with coexisting Weyl points and nodal rings behaves like a Weyl semimetal for its zero Landau energy level and similarly to a nodal ring for its greater-than-zero Landau energy band. In the presence of two scattering potentials: a Gaussian potential and a screened Coulomb potential, we study the thermoelectric transport properties of the system, including both longitudinal and transverse configurations. The longitudinal configuration is the case when the magnetic field, temperature gradient and electric field are all along the $z$-direction, while the transverse configuration chooses the magnetic field to remain along the $z$-direction while the electric field and temperature gradient are along the $x$-direction.

In the longitudinal configuration, the thermoelectric conductivity shows a plateau irrespective of the magnetic field and the Fermi energy for $\delta$-form short-ranged scattering. For long-ranged Gaussian or screened Coulomb potentials, the thermoelectric coefficients strongly depend on the impurity distance or the relative dielectric constant. However, the Seebeck response always increases with the field for various scattering at fixed carrier density or Fermi energy for longitudinal configuration. In the transverse configuration, significant positive magnetoresistance and enhanced thermoelectric conductance are observed for both Gaussian and screened Coulomb scatterings. The Hall conductivity surpasses the longitudinal conductivity, resulting in a Seebeck coefficient that approaches the dissipationless limit with a quadratic increase in the magnetic field, regardless of the nature of scattering. Conversely, the Nernst response exhibits a strong dependence on the specific scattering mechanism. Moreover, the thermoelectric transport properties are notably influenced by the model parameter $D$.


\begin{acknowledgments}
This work was supported by the National Natural Science Foundation of China (Grant No. 11974249).
\end{acknowledgments}

\end{document}